\documentclass[showpacs,twocolumn]{revtex4}

\usepackage{amsmath,amssymb,bm,graphicx}

\begin{document}

\title{Hartree-Fock study of an Anderson metal-insulator transition in the presence of Coulomb interaction: Two types of mobility edges and their multifractal scaling exponents}
\author{ Hyun-Jung Lee$^{1,2}$ and Ki-Seok Kim$^{1}$ }
\affiliation{ $^{1}$Department of Physics, POSTECH, Pohang, Gyeongbuk 790-784, Korea \\ $^{2}$Asia Pacific Center for Theoretical Physics (APCTP), POSTECH, Pohang, Gyeongbuk 790-784, Korea }
\date{\today}

\begin{abstract}
We investigate the role of Coulomb interaction in the multifractality of Anderson metal-insulator transition, where the Coulomb interaction is treated within the Hartree-Fock approximation, but disorder effects are taken into account exactly. An innovative technical aspect in our simulation is to utilize the Ewald-sum technique, which allows us to introduce the long-range nature of the Coulomb interaction into Hartree-Fock self-consistent equations of order parameters more accurately. This numerical simulation reproduces the Altshuler-Aronov correction in a metallic state and the Efros-Shklovskii pseudogap in an insulating phase, where the density of states $\rho (\omega)$ is evaluated in three dimensions. Approaching the quantum critical point of a metal-insulator transition from either the metallic or insulting phase, we find that the density of states is given by $\rho (\omega) \sim |\omega|^{1/2}$, which determines one critical exponent of the McMillan-Shklovskii scaling theory. Our main result is to evaluate the eigenfunction multifractal scaling exponent $\alpha_{q}$, given by the Legendre transformation of the fractal dimension $\tau_{q}$, which characterizes the scaling behavior of the inverse participation ratio with respect to the system size $L$. Our multifractal analysis leads us to identify two kinds of mobility edges, one of which occurs near the Fermi energy and the other of which appears at a high energy, where the density of states at the Fermi energy shows the Coulomb-gap feature. We observe that the multifractal exponent at the high-energy mobility edge remains to be almost identical to that of the Anderson localization transition in the absence of Coulomb interactions. On the other hand, we find that the multifractal exponent near the Fermi energy is more enhanced than that at the high-energy mobility edge, suspected to result from interaction effects. However, both the multifractal exponents do not change even if the strength of the Coulomb interaction varies. We also show that the multifractality singular spectrum can be classified into two categories, confirming the appearance of two types of mobility edges.
\end{abstract}

%\pacs{71.10.Hf, 71.10.-w, 71.10.Fd, 71.30.+h}

\maketitle

\section{Introduction}

Strong fluctuations of eigenfunctions are the characteristic feature of the Anderson metal-insulator transition, which can be quantified by a set of inverse participation ratios, $P_q = \int d^d r |\psi^{2q}({ \bf r})|$ \cite{AMIT_Review}, where $\psi({ \bf r})$ denotes an eigenfunction for a given configuration of disorder. This is nothing but the $(q-1)^{th}$ moment of the probability density $|\psi^{2}({ \bf r})|$, described by $P_q=\int d^d r |\psi^{2(q-1)}({ \bf r})| |\psi^{2}({ \bf r})|$. A disorder average of the inverse participation ratio follows the scaling behavior of $\langle P_q\rangle \propto L^{-d(q-1)}$ in a metallic phase, where the eigenfunction at the Fermi energy is extended. Here, $L$ is the size of a system and $d$ is its dimension. On the other hand, the eigenfunction is localized in an Anderson insulating state, and the disorder average of the inverse participation ratio becomes independent of the system size, given by $\langle P_q\rangle \propto L^0$. In the vicinity of the Anderson metal-insulator transition, the inverse participation ratio shows an anomalous scaling behavior with respect to the system size $L$, given by $\langle P_q\rangle \propto L^{-\tau_q}$ with $\tau_q=D_q(q-1)$, where $D_q \not= d$ is the fractal dimension for each moment. If one replaces the probability density of an eigenfunction with an order parameter, he/she can calculate quantum mechanical averages for multiple moments of the order parameter. It turns out that such higher moments do not show critical scaling behaviors in conventional continuous quantum phase transitions. On the other hand, all the moments of eigenfunctions give rise to fractal behaviors in the vicinity of the Anderson metal-insulator transition, referred to as multifractality and regarded to be an essential feature of the Anderson metal-insulator transition \cite{Comments_Multifractality}.

Nature of the eigenfunction multifractality has been discussed both intensively and extensively in the vicinity of the Anderson metal-insulator transition. Analytical calculations based on the nonlinear $\sigma-$model field theory, which describes effective interactions between diffusions and Cooperons, turn out to be consistent with essentially exact numerical studies for fractal dimensions \cite{Multifractality_NLsM_vs_Numerics}. A natural question would be on the role of electron correlations in the multifractality of the Anderson metal-insulator transition. Recently, tunneling experiments on Ga$_{1-x}$Mn$_{x}$As have shown that the nature of eigenfunction multifractal correlations in the vicinity of the metal-insulator transition differs from that without electron correlations \cite{Multifractality_Interaction_Experiment}, suggesting that not only the eigenfunction multifractality survives electron interactions but also its nature gets modified. Motivated from such tunneling measurements, the fractal dimension of $D_{q}$ has been evaluated not only based on the nonlinear $\sigma-$model approach in the presence of Coulomb interaction \cite{Multifractality_Interaction_NLsM}, but also based on a numerical study, where the Coulomb interaction is treated within the Hartree-Fock approximation, but disorder effects are taken into account exactly \cite{Multifractality_Interaction_Numerics}. However, the nature of eigenfunction multifractality in the presence of Coulomb interactions is not still well understood, being under current debates.
%
%(Un)fortunately, their values turn out to be not much close, which requires further studies.
%

We reexamine the effect of Coulomb interaction on the multifractality of Anderson metal-insulator transition, resorting to the Hartree-Fock approximation for the Coulomb interaction, where disorder effects are carried out exactly. Our major technical innovation is to utilize the Ewald-sum technique \cite{Ewald_Sum_Technique1,Ewald_Sum_Technique2,Ewald_Sum_Technique3}, which allows us to introduce the long-range nature of the Coulomb interaction into Hartree-Fock self-consistent equations of order parameters rather accurately. Based on this improved numerical technique, we evaluate the multifractal scaling exponent $\alpha_{q}$, given by the Legendre transformation of the fractal dimension $\tau_{q}$ discussed above, which characterizes the eigenfunction multifractal nature near the metal-insulator transition. Here, we focus on a characteristic disorder strength slightly below a critical value of the Anderson metal-insulator transition in three dimensions, above which all quantum states become localized. As a result, electrons at the Fermi energy remain delocalized to show diffusive dynamics in the absence of electron correlations. On the other hand, electrons at high energies become localized, where the density of states are much smaller than that at the Fermi energy and disorder potentials are not screened sufficiently due to the lack of the density of states. The characteristic energy for the Anderson localization is called the mobility edge \cite{Anderson_Localization_Review}. The multifractal nature of the mobility edge has been well understood in the absence of Coulomb interaction as discussed above.

Introducing the Coulomb interaction into the diffusive metallic phase at the characteristic disorder strength below the Anderson localization, the density of states at the Fermi energy evolves to be suppressed due to the Altshuler-Aronov correction \cite{Altshuer_Aronov_Correction} in the case of weak Coulomb interactions. Increasing the Coulomb interaction further, the density of states at the Fermi energy vanishes to show the Efros-Shklovskii pseudogap feature \cite{Coulomb_Gap_Review}. Our numerical analysis confirms the emergence of a mobility edge in the vicinity of the Coulomb-gap formation due to the suppression of the density of states in addition to the high-energy mobility edge involved with the Anderson localization without electron correlations. The emergence of the mobility edge near the Fermi energy seems to be consistent with the observation of the recent tunneling experiment \cite{Multifractality_Interaction_Experiment} although this measurement does not identify the mobility edge at a high energy. We find that the multifractal exponent at the high-energy mobility edge remains to be almost identical to that in the absence of Coulomb interactions. On the other hand, we reveal that the multifractal exponent near the Fermi energy is more enhanced than that at the high-energy mobility edge, suspected to result from interaction effects. However, both the multifractal exponents do not change even if the strength of the Coulomb interaction varies. We also show that the multifractality singular spectrum can be classified into two categories, confirming the appearance of two types of mobility edges.

Before going further, we would like to introduce two recent studies investigating the role of Coulomb interactions in the Anderson metal-insulator transition \cite{Coulomb_Disorder_DFT_I,Coulomb_Disorder_DFT_II}. Although these two studies are based on the density functional theory approximation, which differs from that of the present study, both papers pointed out that the Coulomb interaction changes the universality class of the Anderson metal-insulator transition, resorting to the multifractal analysis. In particular, Ref. \cite{Coulomb_Disorder_DFT_II} suggested a possible resolution of the long-standing ``exponent puzzle" due to the interplay between conduction and impurity states.

\section{Hartree-Fock approximation and Ewald summation technique}

We start from a disordered Hubbard Hamiltonian of spinless fermions on a three-dimensional cubic lattice of the size $L^{3}$, given by
\begin{equation}
  H=\sum_{\langle ij\rangle }(-t_{ij}+\varepsilon_i \delta_{ij})c_i^\dagger c_j+\frac{1}{2}\sum_{ij}U_{ij}\delta n_i\delta n_j .
  \label{modelHam}
\end{equation}
Here, the Coulomb interaction of $U_{ij}=\frac{e^2}{\kappa r_{ij}}$ is taken into account, where $\delta n_i=n_i-K$ is the fluctuation of the electron occupation $n_i$ around the mean value $K$. $e^{2} / \kappa$ with a dielectric constant $\kappa$ is referred to as the strength of the Coulomb interaction, denoted by $U$. $t_{ij}$ is a parameter for nearest neighbor hopping, set to be $t=1$ as the unit of energy. Onsite energies $\varepsilon_i$ are random, independently and uniformly distributed in $\varepsilon_i \in \left[ -W,W\right]$. The chemical potential $\mu$ can be renormalized by interactions, but chosen so as to keep the average density $K = 1/2$. For non-interacting particles at $U=0$, the Anderson metal-insulator transition occurs at a critical disorder strength $W_c=8.25$ \cite{AMIT_Critical_Disorder_Strength}, where the mobility edge comes from a high energy to the Fermi energy, localizing all quantum states of electrons.

We attack this problem numerically based on the Hartree-Fock approximation. Following Ref. \cite{Multifractality_Interaction_Numerics}, we write down an effective single-particle model with self-consistent onsite energies and hopping amplitudes
\begin{equation}
    H_{HF}=\sum_i \tilde{V}_i c_i^\dagger c_i -\sum_{ij} \tilde{t}_{ij} c_i^\dagger c_j +h.c.
    \label{H_nonint}
\end{equation}
Here, the self-consistent onsite potential energy and the self-consistent hopping kinetic energy are given by
\begin{eqnarray}
  \tilde{V}_i&=&\varepsilon_i+\sum_{j}\frac{U}{\bf |r_i-r_j |} \left[\langle c_j^\dagger c_j\rangle-K\right]-\mu , \label{selfconsol1} \\
  \tilde{t}_{ij}&=&t_{ij}+\frac{U}{\bf |{r_i-r_j}|}\langle c_j^\dagger c_i\rangle , \label{selfconsol2}
\end{eqnarray}
respectively, where $\langle ... \rangle$ denotes an ensemble average for a given disorder configuration. The effective onsite energy $\tilde{V}_j$ is renormalized by the interaction-induced Hartree term, which leads to ``correlated" on-site energies. The effective hopping parameter $\tilde{t}_{ij}$ is renormalized by the Fock term, where long-range hopping processes are generated by the Coulomb interaction. Here, we carry out exact diagonalization on a cubic three-dimensional lattice of the linear size $L=10$ $\sim$ $L = 24$.

A set of parameters to be determined self-consistently contains the ensemble average of all-range hopping $\langle c_i^\dagger c_j\rangle$ including the local density $\langle c_i^\dagger c_i\rangle$. In order to find a self-consistent solution, we begin with a random initial guess for all parameters, which should satisfy the condition
\begin{equation}
  \sum_i \langle n_i \rangle = \mathcal{N}_e,
\end{equation}
where the number of particles $\mathcal{N}_e=N/2$ is fixed at half filling in our simulation. Based on this initial condition, we diagonalize the effective Hamiltonian and find eigenfunctions $\psi_m({\bf r})$ and eigenvalues $\varepsilon_m$ for a given disorder configuration. Then, we obtain the ensemble average of all-range hopping $\langle c_i^\dagger c_j\rangle$ and the local density $\langle c_i^\dagger c_i\rangle$, resorting to
\begin{eqnarray}
  \langle n_i\rangle &=&\sum_m |\psi_m({\bf r}_i)|^2 f(\varepsilon_m), \\
  \langle c_i^\dagger c_j\rangle&=& \sum_m \psi_m^*({\bf r}_i) \psi_m({\bf r}_j)  f(\varepsilon_m),
\end{eqnarray}
regarded to be defining equations, where $f(\varepsilon_m)$ is the Fermi-Dirac distribution function with the chemical potential $\mu$. The chemical potential is adjusted to assure the system at half filling, determined by
\begin{equation}
  \mathcal{N}_e= \sum_m f(\varepsilon_m).
\end{equation}
Inserting these parameters into both Hartree-Fock self-consistent Eqs.~(\ref{selfconsol1}) and ~(\ref{selfconsol2}), we obtain an updated set of parameters, $\tilde{V}_{i}$ and $\tilde{t}_{ij}$, where $\tilde{t}_{ij}$ becomes long ranged. These renormalized parameters are introduced into the effective Hartree-Fock Hamiltonian Eq. (\ref{H_nonint}). We diagonalize the updated Hartree-Fock Hamiltonian and perform this iteration procedure until the output set of parameters converges within $10^{-4}$ of uncertainty. After convergence, the final set of eigenfunctions $\psi_m({\bf r})$ and eigenvalues $\varepsilon_m$ is used to compute physical quantities such as density of states and correlation functions. The result is then averaged over an ensemble of various disorder realizations, which are randomly selected by the rectangular distribution of a disordered potential $\varepsilon_i \in \left[ -W,W\right]$.

An essential point in solving these Hartree-Fock self-consistent equations is how to deal with the long-range nature of Coulomb interactions. In order to clarify the role of long-range interactions in the Hartree-Fock approximation, we implement the Ewald summation technique, where the Hartree term is split into two parts: a real-space portion based on a short-range interaction potential whose pairwise sum converges quickly and a long-range portion based on a slowly-varying interaction potential whose pairwise sum converges relatively quickly in a reciprocal space \cite{Ewald_Sum_Technique1,Ewald_Sum_Technique2,Ewald_Sum_Technique3}. The optimal implementation of the Ewald technique allows us to resolve the long-standing issue of the ill-convergence of the long-range potential and the multiplicity of Hartree-Fock solutions near the Anderson-Mott transition, which has been also reported in a recent Hartree-Fock numerical study \cite{Multifractality_Interaction_Numerics}. We refer all details, certainly important, to Appendix.

Based on the Ewald summation technique, the typical number of iterations required for a convergent solution for one realization of disorder is $\sim 30$. The bottleneck process in the overall computing steps is the $ \sim L^9$ operations to calculate the effective hopping matrix with the size $L^3 \times L^3$, which is performed in parallel using MPI. For a single disorder realization, the total time at $L=18 (L=24)$ is of the order of 50 hours (60 days) $/(\#$cores) for each parameter set of interaction and disorder strengths. For $50 (10)$ different disorder realizations, the total time at $L=18 (L=24)$ is of the order of 10 days (2 months) if $9(12)$ cores for parallel computing are used.
\vspace{-0.3cm}

\section{Result and Discussion}
\subsection{Density of states}
\begin{figure}
\begin{center}
  \vspace{0.cm}
  \begin{tabular}{c}
    \includegraphics[scale=0.325]{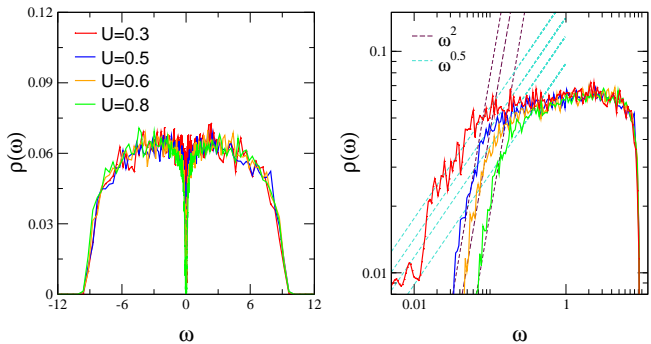}
    \end{tabular}
\end{center}
\vspace{-0.5cm}
\caption{Density of states for various interaction parameters of $0.3 \leq U \leq 0.8$ at $W = 7$ for a cubic lattice with the size $N=L^3=18^3$. Here, the interaction strength $U$ is given by the Coulomb interaction $V_{C}(|\bm{r}_{i} - \bm{r}_{j}|) = \frac{U}{|\bm{r}_{i} - \bm{r}_{j}|}$ and the critical disorder strength is $W_{c} \sim 8.25$ for the Anderson metal-insulator transition in three dimensions. The left panel displays a conventional plot for the density of states, which reproduces the Altshuler-Aronov correction in a metallic state and the Coulomb-gap feature in an insulating phase, clarified in the log-log plot of the right panel \cite{IRcutoff_mismatch}. An essential point is that the scaling behavior of $\rho(\omega) \sim |\omega|^{1/2}$, the Altshuler-Aronov correction in three dimensions, continues to dominate, approaching the metal-insulator transition from the metallic phase. This suggests that the critical interaction strength $U_{c}$ for the metal-insulator transition is estimated to be $0.3 < U_{c} < 0.5$ at $W = 7$, where the $\rho(\omega) \sim |\omega|^{1/2}$ scaling behavior turns into $\rho(\omega) \sim |\omega|^{2}$. This quantum critical regime shrinks in the insulating state and the Coulomb-gap scaling of $\rho(\omega) \sim |\omega|^{2}$ dominates in this region.}
\label{dos1}
\end{figure}

We show the density of states $\rho(\omega)$ in Fig. 1. Here, the interaction strength $U$ in the Coulomb potential $V_{C}(|\bm{r}_{i} - \bm{r}_{j}|) = \frac{U}{|\bm{r}_{i} - \bm{r}_{j}|}$ is varied from $U = 0.3$ to $U = 0.8$, and the disorder strength is fixed to be $W = 7$ below the critical disorder strength $W_{c} \sim 8.25$ \cite{AMIT_Critical_Disorder_Strength} for the Anderson metal-insulator transition in three dimensions. When the interaction strength is less than a critical value $U_{c}$, the density of states remains to be finite at the Fermi energy, but gets suppressed due to interaction corrections. This suppression is referred to as the Altshuler-Aronov correction, where diffusive electrons acquire strong renormalization effects even in the Hartree-Fock level \cite{Altshuer_Aronov_Correction}. The right panel confirms the typical suppression behavior of $\rho(\omega) - \rho_{0} \sim |\omega|^{1/2}$ in thee dimensions, where $\rho_{0}$ is the suppressed density of states at the Fermi energy \cite{IRcutoff_mismatch}. An interesting point is that the frequency scaling behavior of the Altshuler-Aronov type correction persists up around the critical point of a metal-insulator transition. This scaling behavior near the metal-insulator transition should be distinguished from the Altshuler-Aronov correction in the weak coupling approach, given by small corrections in the density of states. Here, the density of states changes more than two times, which seems to be beyond the weak coupling approach. This determines one critical exponent of the McMillan-Shklovskii scaling theory \cite{McMillan_Shklovskii_scaling_theory}. We point out that the critical exponent of $\rho(\omega) - \rho_{0} \sim |\omega|^{1/2}$ is consistent with that of the recent numerical study \cite{Multifractality_Interaction_Numerics}.
\begin{figure}
\begin{center}
  \vspace{0.cm}
  \begin{tabular}{c}
    \includegraphics[scale=0.3]{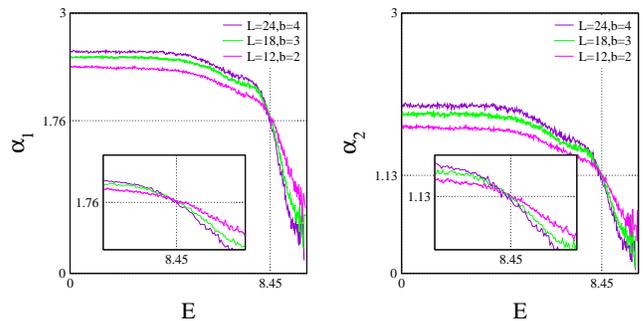}
    \end{tabular}
\end{center}
\vspace{-0.5cm}
\caption{Multifractal scaling exponents ($\alpha_1$ and $\alpha_2$) for various system sizes in the non-interacting case ($U=0$) with a disorder strength $W=7$ slightly below the critical disorder strength $W_{c} \sim 8.25$ of the Anderson metal-insulator transition. Here, $L$ denotes the size of a system and $b$ represents the size of a block, where the ratio of $(L/b)^{3} = 6^{3}$ is fixed. Enhancing the size of a system, the multifractal scaling exponent increases (decreases) in the metallic (insulating) region, where the energy is less (larger) than the mobility edge, i.e., $E < E_{m}$ ($E > E_{m}$). It does not depend on the system size at the mobility edge, identified with the critical energy for the Anderson-metal insulator transition. Based on this physics, we can determine not only the multifractal scaling exponents as $\alpha_{1} = 1.76$ and $\alpha_{2} = 1.13$ but also the position of the mobility edge as $E_{m} = 8.45$, all of which are consistent with previous studies \cite{AMIT_Review}.}
\vspace{-0.5cm}
\label{alpha_U0_W7}
\end{figure}

When the interaction parameter exceeds the critical value, the density of states at the Fermi energy vanishes, given by $\rho(\omega) \sim |\omega|^{2}$ and identified with the Efros-Shklovskii pseudogap \cite{Coulomb_Gap_Review}. This Coulomb gap feature starts to appear around $U = 0.5$ and becomes almost completed around $U = 0.8$, clarified by the right panel. Interestingly, the McMillan-Shklovskii scaling coexists with this Coulomb-gap scaling at $U = 0.5$. This evolution implies that the critical value of the interaction parameter is around $0.3 < U_{c} < 0.5$, which identifies the metal-insulator transition, where the density of states vanishes at the Fermi energy.
\vspace{-0.6cm}
\subsection{Multifractal analysis at the mobility edge of the Anderson model in the absence of electron correlations}
\begin{figure}
\begin{center}
  \vspace{0.cm}
  \begin{tabular}{cc}
    \includegraphics[scale=0.3]{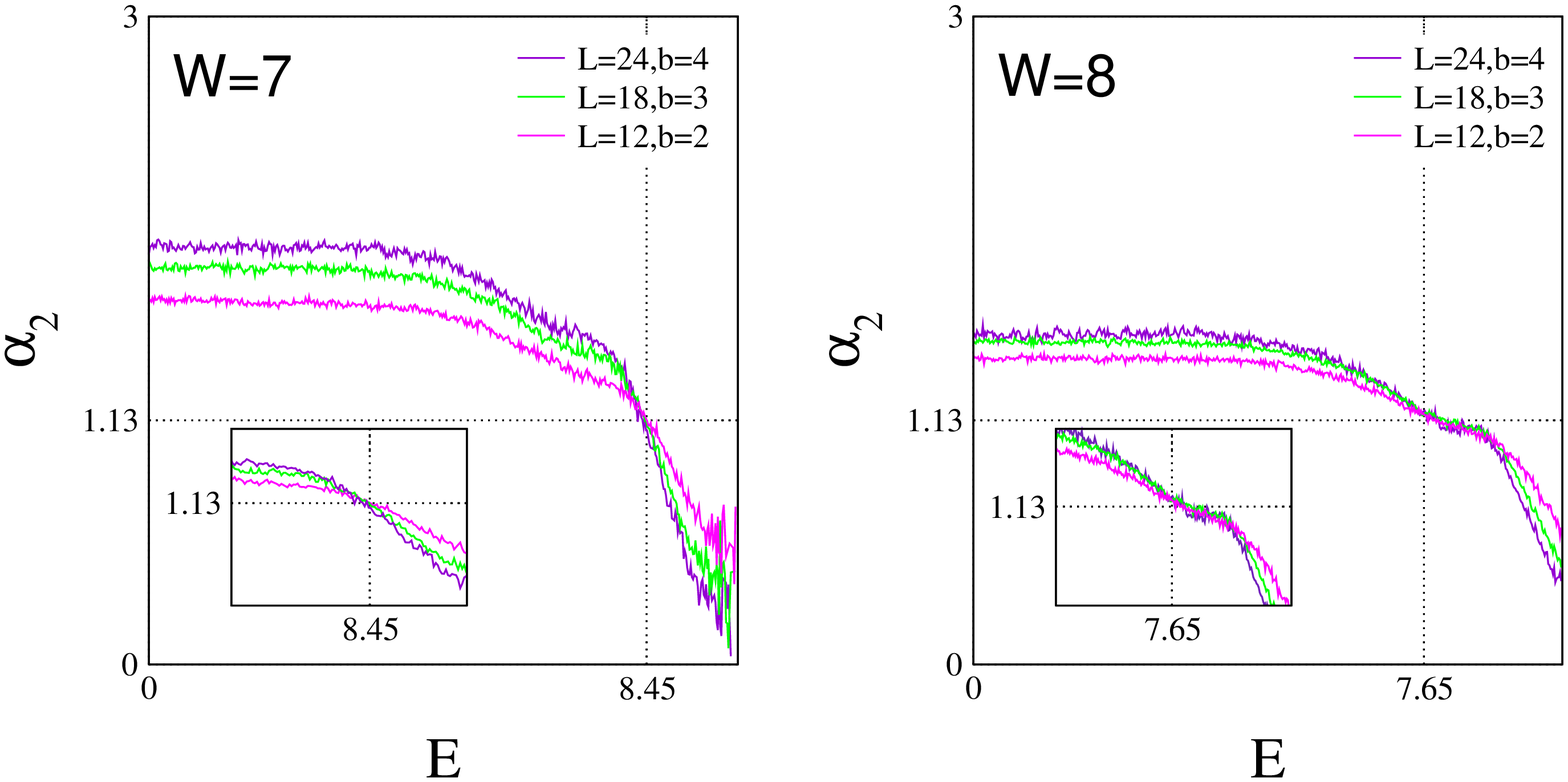}
    \end{tabular}
\end{center}
\vspace{-0.5cm}
\caption{Multifractal scaling exponent $\alpha_2$ for two disorder strengths $W=7$ and $W=8$. Near the critical disorder strength ($W=8 \sim W_c = 8.25$), the  mobility edge, where the scale-invariance is realized, gets broader ($7.6 < E < 8.8$). The multifractal scaling exponent does not depend on the disorder strength $W$ at the mobility edge.}
\vspace{-0.3cm}
\label{alpha_U0_W7_W8}
\end{figure}
In order to find the fractal dimension $\tau_{q} = D_{q} (q-1)$ numerically, it is more convenient to introduce a coarse graining box with a volume of $b^{d}$, where the unit of an eigenfunction intensity is given by \cite{f_alpha}
\begin{equation}
  \mu_{k}(b) \equiv \sum_{j \in \mbox{box}_{k}} |\psi_{j}^{2}| .
\label{boxprob}
\end{equation}
Here, $j$ is a lattice site and box$_{k}$ is the coarse graining box with an effective index $k$. Then, moments of eigenfunction intensities are naturally introduced in the following way
\begin{equation}
  P_{q}(b) \equiv \sum_{k} [\mu_{k}(b)]^{q} .
\end{equation}
Accordingly, the fractal dimension can be defined as
\begin{equation}
  \tau_q = \lim_{\lambda\rightarrow 0}\frac{\ln \langle P_q \rangle}{\ln \lambda} ,
  \label{tauq}
\end{equation}
where $\lambda=b/L$.

The disorder average of the inverse participation ratio can be reformulated with the introduction of the distribution function $\mathcal{P}(|\psi^{2}|)$ for eigenfunction intensities, given by \cite{AMIT_Review}
\begin{equation}
  \mathcal{P}(|\psi^{2}|) \sim \frac{1}{|\psi^{2}|} L^{-d + f(- \ln |\psi^{2}| / \ln L)} ,
  \label{Distribution_Function_Alpha}
\end{equation}
where the function of $f(- \ln |\psi^{2}| / \ln L)$ defines the distribution function, referred to as the multifractal singularity spectrum. Actually, the disorder average of the moments $|\psi^{2q}|$ is expressed as
\begin{eqnarray}
  \langle P_q \rangle \equiv L^{d} \langle |\psi^{2q}| \rangle &=& L^{d} \int d |\psi^{2}| |\psi^{2 q}| \mathcal{P}(|\psi^{2}|) \nonumber \\ & \sim & \int d \alpha L^{- q \alpha + f(\alpha)} ,
  \label{Averaged_IPR}
\end{eqnarray}
where $\alpha = - \ln |\psi^{2}| / \ln L$ was introduced. Taking into account the limit of large $L$, this integral can be performed in the saddle-point approximation, resulting in
\begin{equation}
  \tau_{q} = q \alpha_{q} - f(\alpha_{q})
\label{tau_q2}
\end{equation}
with $q = \frac{d f(\alpha_{q})}{d \alpha_{q}}$ and $\alpha_{q} = \frac{d \tau_{q}}{d q}$. Here, we focus on the multifractal scaling exponent $\alpha_{q}$, which results from the Legendre transformation of the fractal dimension $\tau_{q}$ \cite{f_alpha}. Investigating the scaling behavior of such exponents with respect to the system size, we can determine the mobility edge and the multifractal exponent reliably.
\begin{figure}
\begin{center}
  \vspace{0.cm}
  \begin{tabular}{cc}
    \includegraphics[scale=0.3]{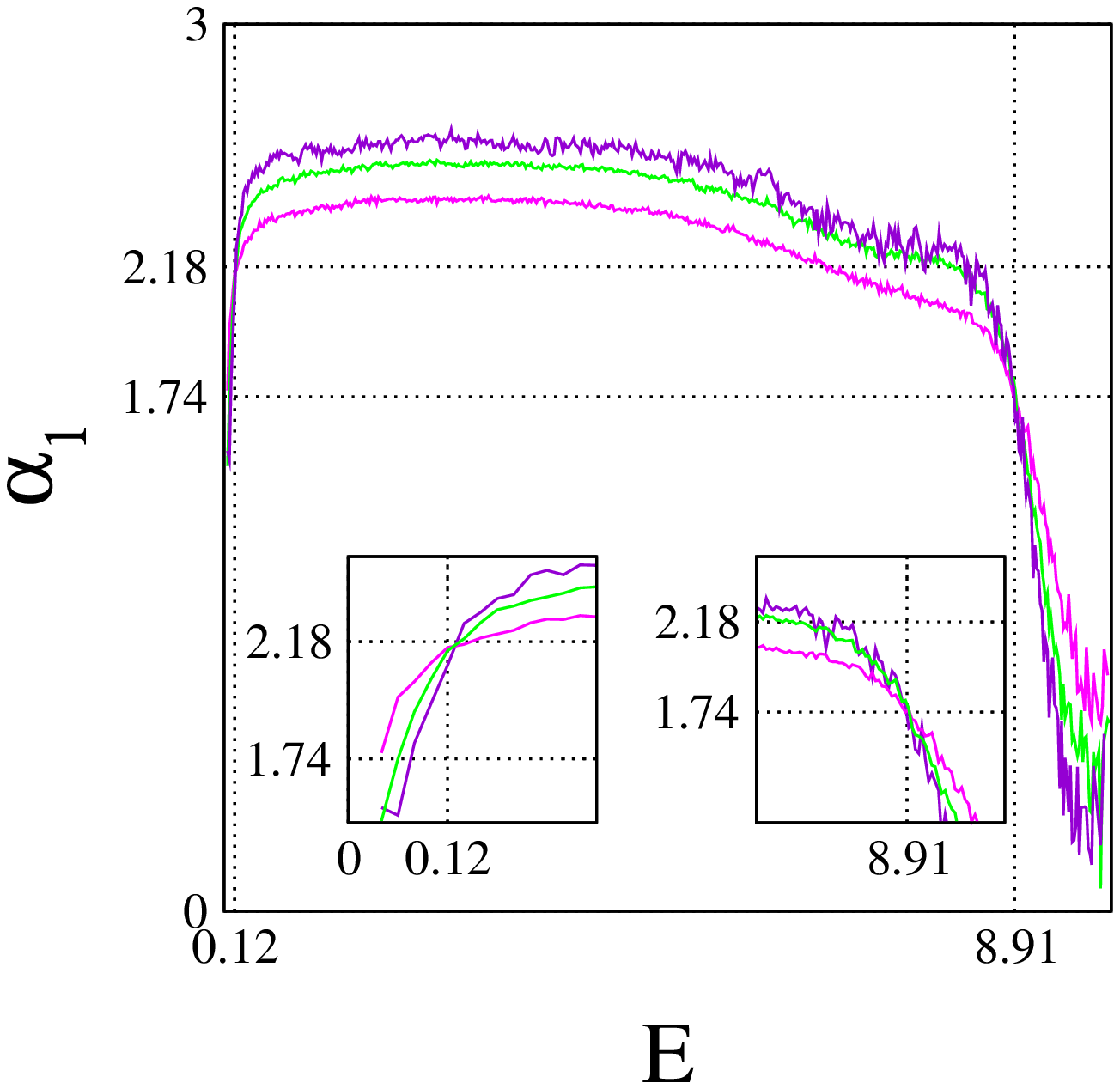} &
    \includegraphics[scale=0.3]{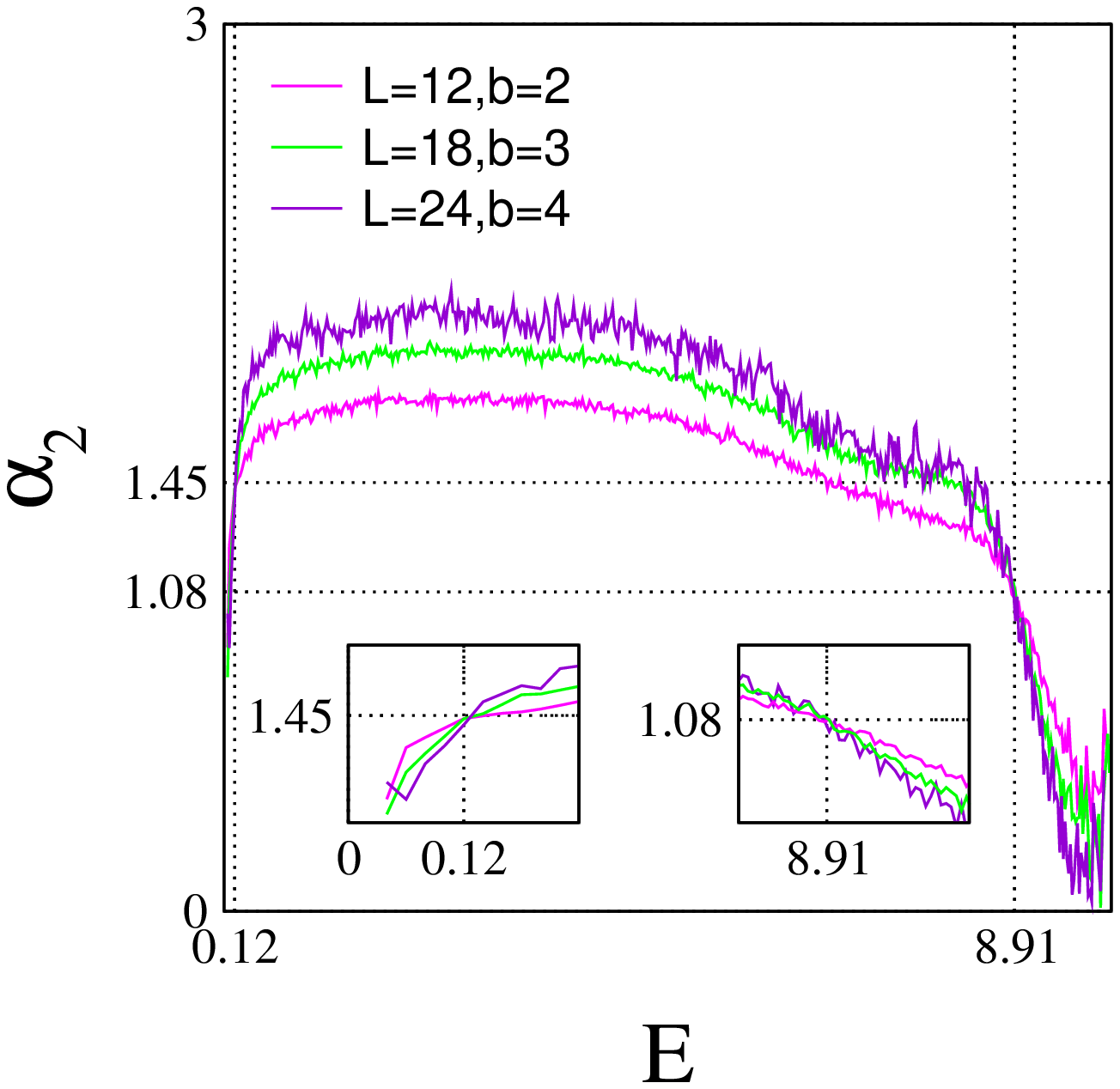} \\
    \vspace{-0.3cm}\\
    \includegraphics[scale=0.3]{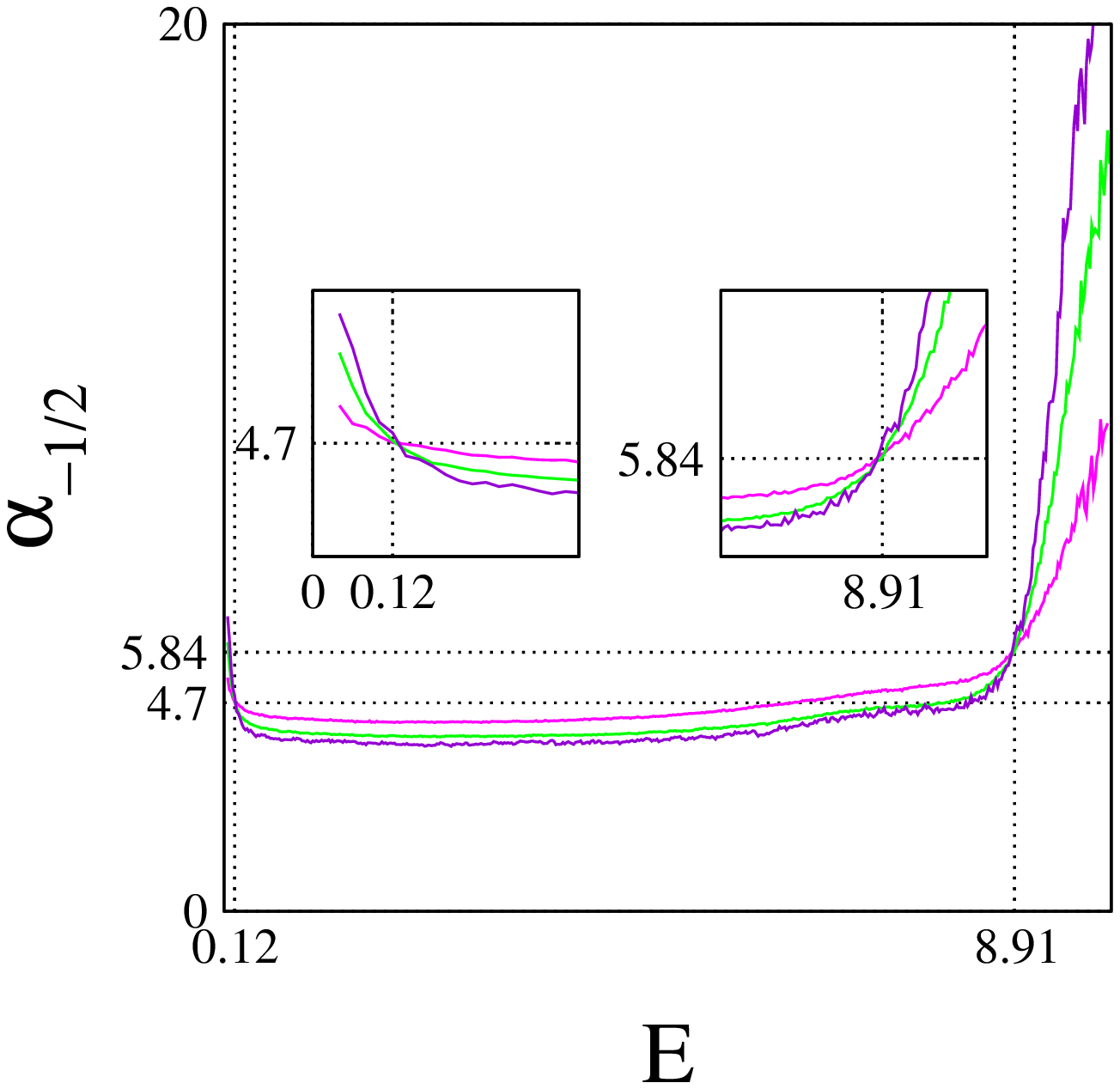} &
    \includegraphics[scale=0.3]{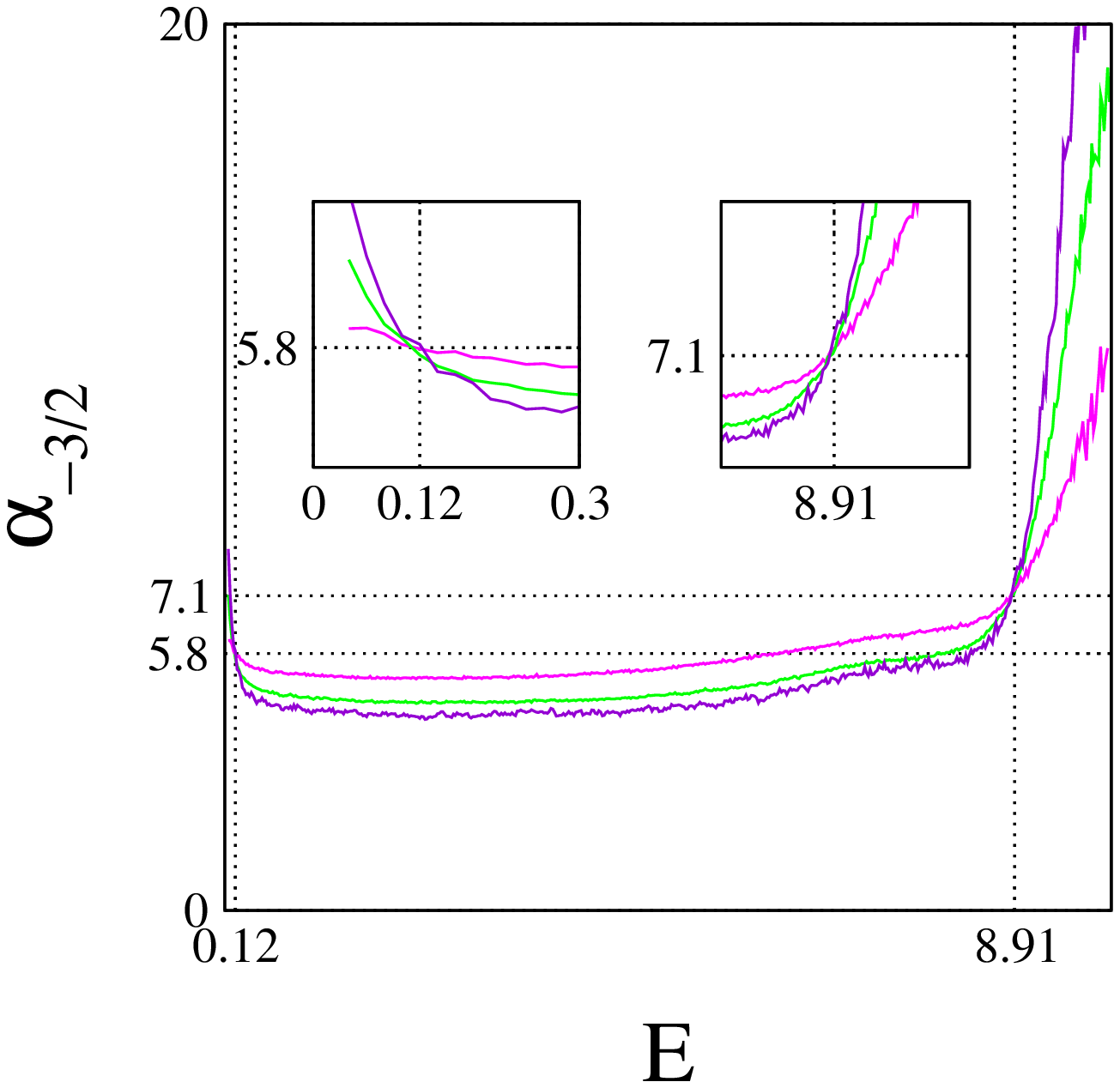}
    \vspace{-0.cm}
    \end{tabular}
\caption{Multifractal scaling exponents of $\alpha_1$, $\alpha_2$, $\alpha_{-0.5}$, and $\alpha_{-1.5}$ in the presence of the Coulomb interaction with $U=0.8$ at the disorder strength $W=7$. All these figures confirm the emergence of two types of mobility edges, one of which occurs near the Fermi energy and the other of which appears at a high energy, where the density of states at the Fermi energy shows the Coulomb-gap feature. It turns out that the multifractal scaling exponents at the high-energy mobility edge remain identical to those in the absence of Coulomb interactions. On the other hand, the low-energy mobility edge results from electron correlations, and the multifractal scaling exponents at the low-energy mobility edge differ from those at the high-energy mobility edge.}
\label{alpha_qvar}
\end{center}
\vspace{-0.5cm}
\end{figure}

\begin{figure}
\begin{center}
  \vspace{0.cm}
  \begin{tabular}{cc}
    \includegraphics[scale=0.3]{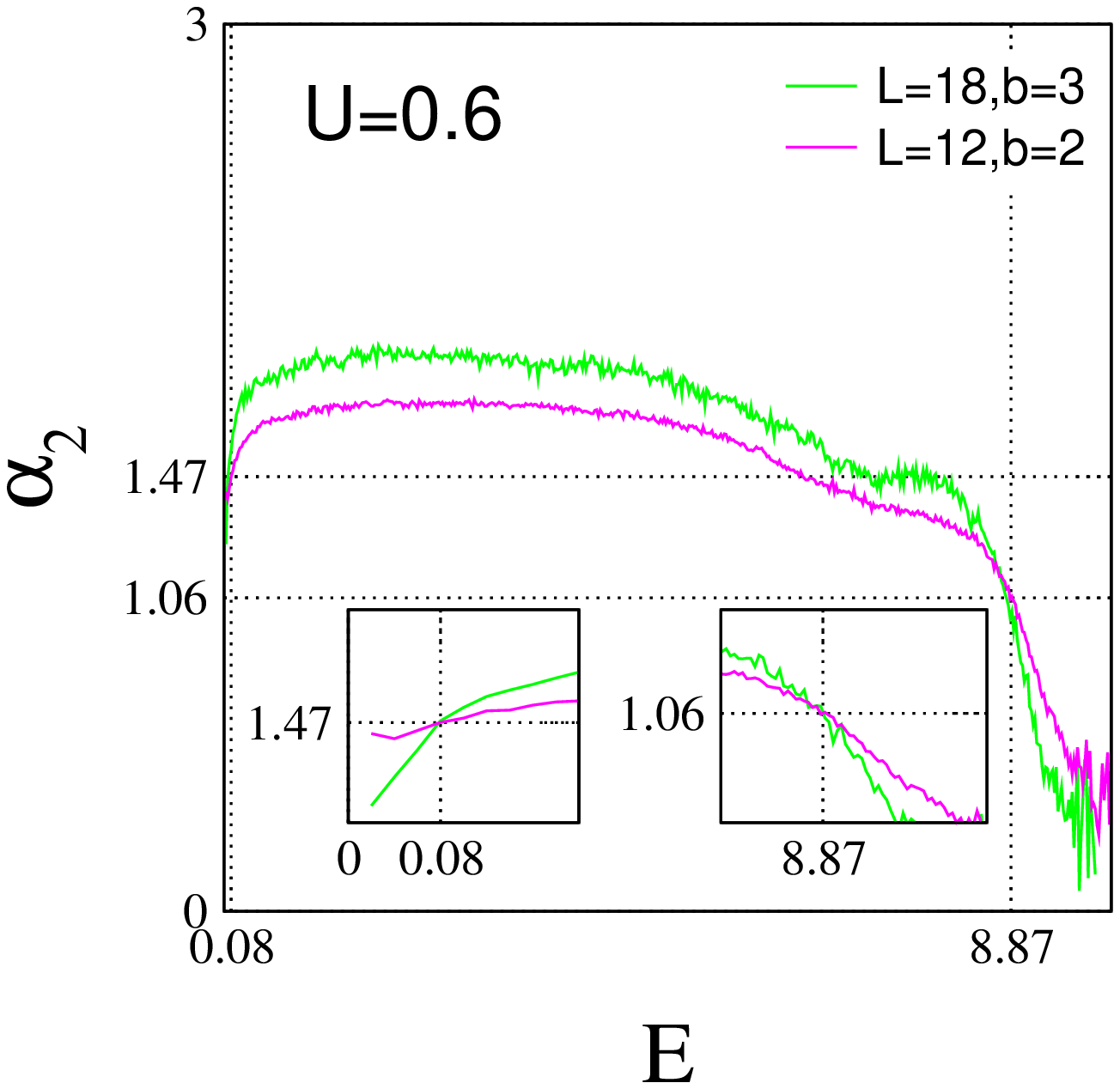} &
    \includegraphics[scale=0.3]{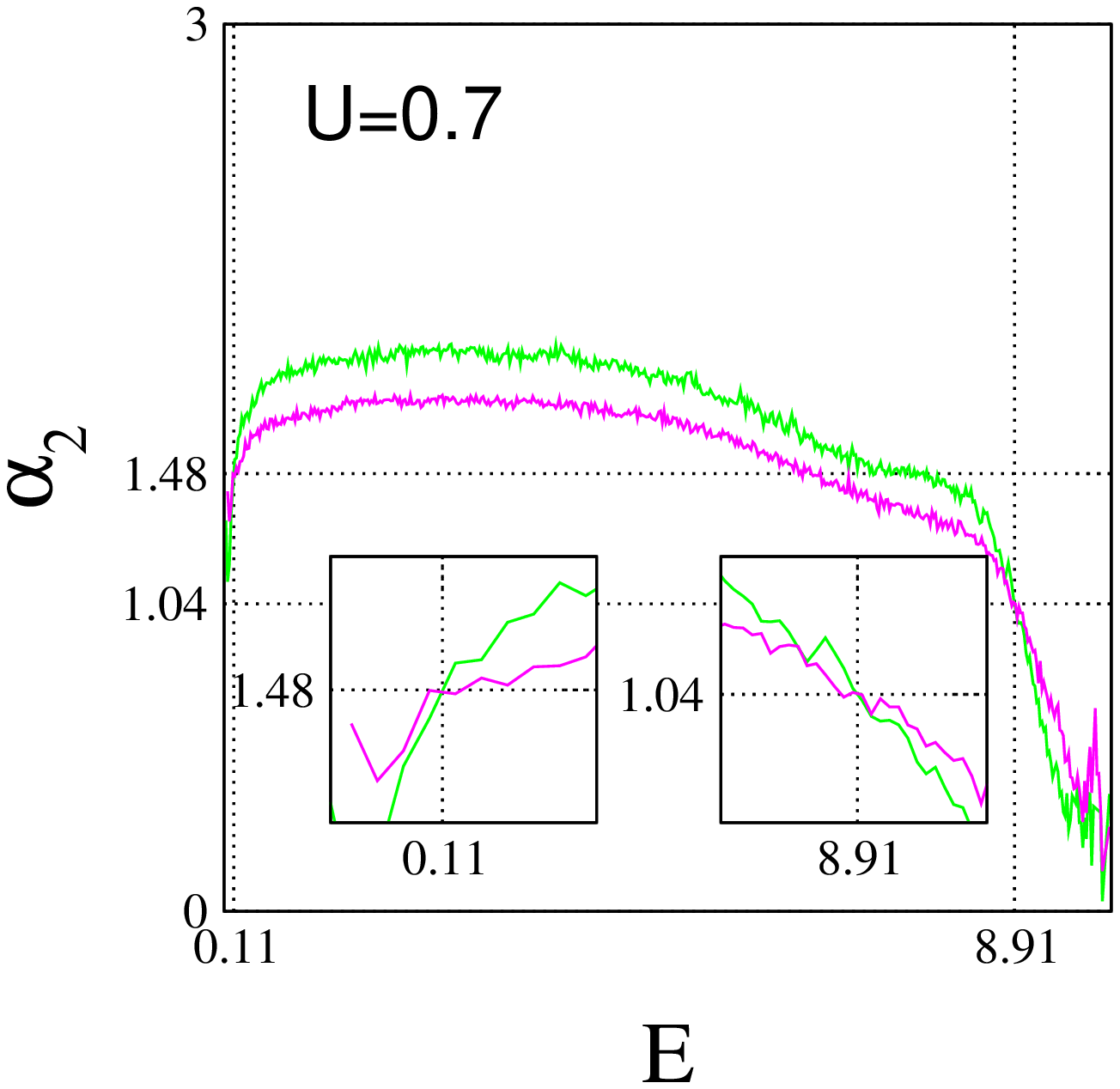} \nonumber \\
    \vspace{-0.3cm}\\
    \includegraphics[scale=0.3]{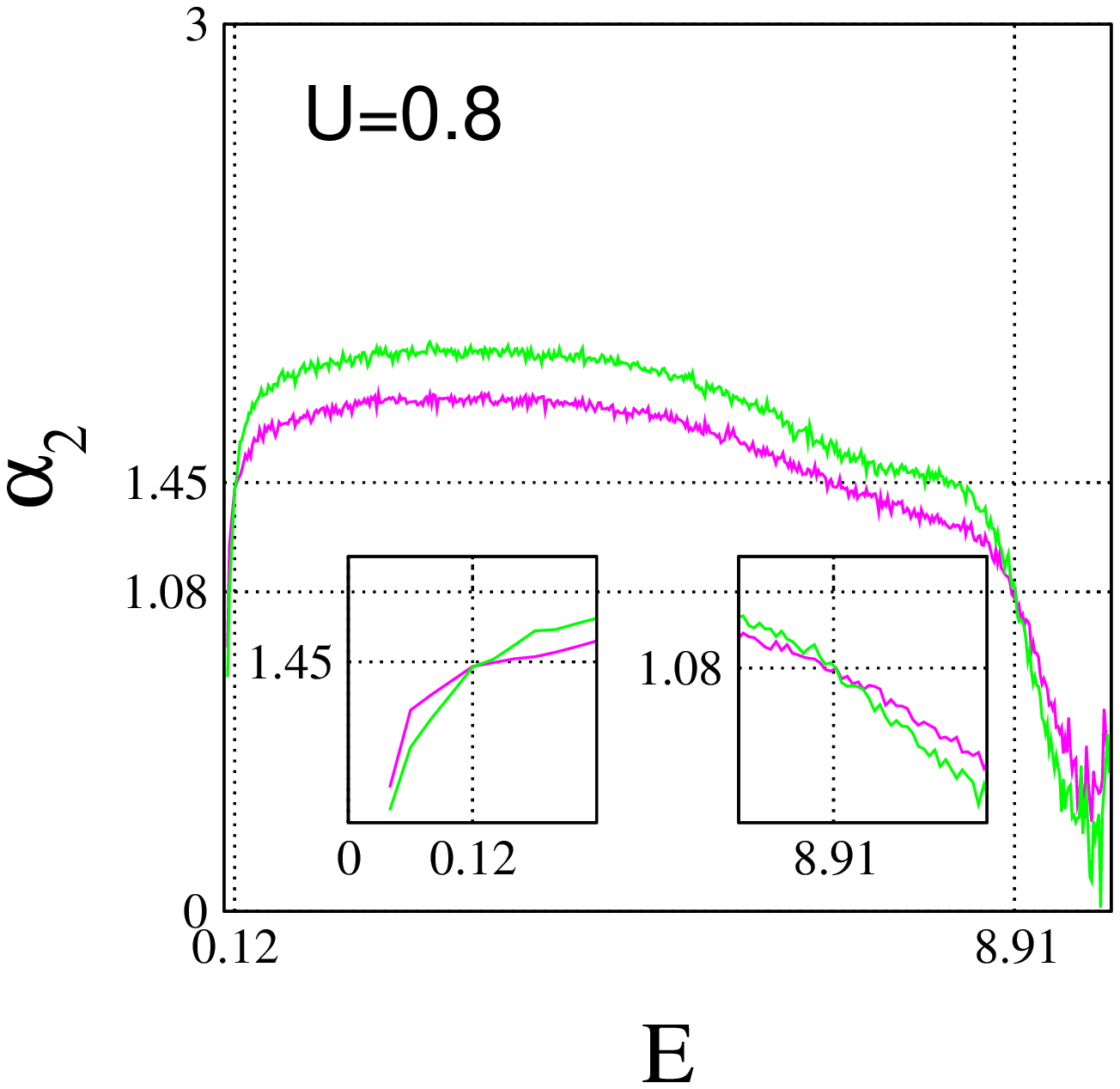} &
    \includegraphics[scale=0.3]{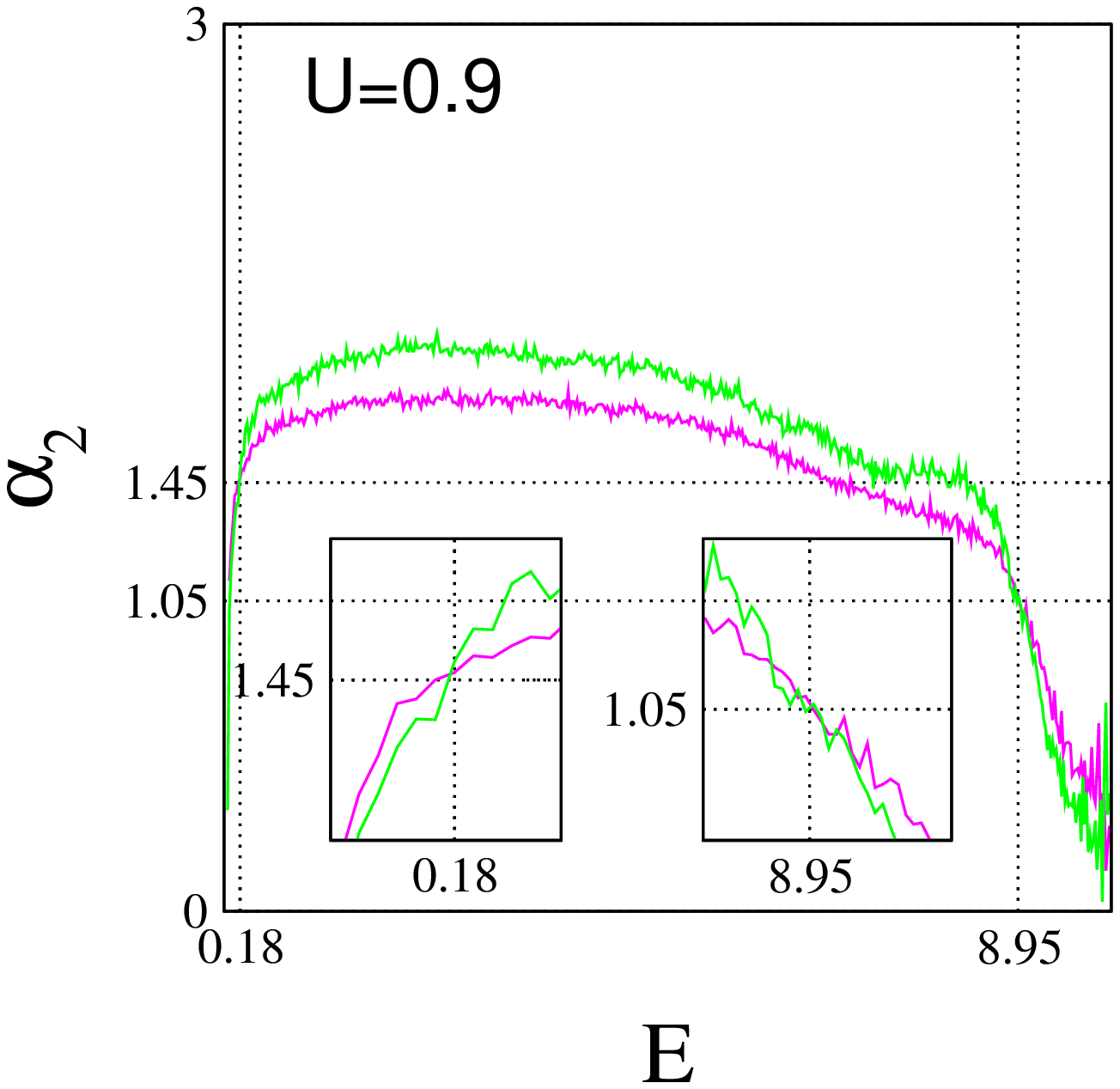} \nonumber \\
    \end{tabular}
    \vspace{-0.cm}
\caption{Multifractal scaling exponent $\alpha_2$ for various interaction parameters of $U=0.6$, $U = 0.7$, $U=0.8$, and $U = 0.9$ at the disorder strength $W=7$. The multifractal scaling exponent not only at the high-energy mobility edge but also at the low-energy mobility edge does not depend on the Coulomb interaction. This result seems to be consistent with a non-linear $\sigma-$model study although the value itself differs from our numerical value \cite{Multifractality_Interaction_NLsM}.}
\label{alpha_Uvar}
\end{center}
\vspace{-0.3cm}
\end{figure}

First, we show multifractal scaling exponents of $\alpha_1$ and $\alpha_2$ for various system sizes in the non-interacting case ($U=0$) with a disorder strength $W=7$ slightly below the critical disorder strength $W_{c} \sim 8.25$ of the Anderson metal-insulator transition, given by Fig.~\ref{alpha_U0_W7}. The horizontal axis is the energy scale, where the zero point is defined as the center of a band. The vertical axis corresponds to the multifractal scaling dimension $\alpha_q$. These critical exponents are evaluated for three different system sizes of $L = 12$, $L = 18$, and $L = 24$ with a fixed value of $\lambda=1/6$. It is clear that there exists a crossing point, denoted by $E_{m} = 8.45$, referred to as the mobility edge, where the multifractal scaling exponent $\alpha_q$ exhibits scale-invariance irrespective of the system size $L$ and the size of a coarse graining box. In the region of $E<E_m$, the multifractal scaling exponent $\alpha_q$ increases as the system size $L$ grows, regarded to be a characteristic feature of an electronic wave function extended over a space. In the clean limit, i.e., the absence of impurity scattering ($W=0$), $\alpha_q$ is proportional to the value of the spatial dimension of a system, given by $\alpha_q = d(q-1)$. See Fig.~\ref{alpha_U0_W7_W8}. In the presence of disorder scattering ($W \neq 0$), $\alpha_q$ shows a non-linear $q$ dependence, where $\alpha_q/(q-1)$ is always less than the spatial dimensionality $d$. In the region of $E>E_m$, the multifractal scaling exponent $\alpha_q$ decreases as $L$ increases, which indicates that the electronic wave function is confined within a finite volume.

Next, we discuss the evolution of the multifractal scaling exponent $\alpha_{2}$ with respect of the disorder strength $W$, shown in Fig.~\ref{alpha_U0_W7_W8}. The multifractal scaling exponent $\alpha_{2}$ decreases with increasing disorder strength $W$, which indicates the progress of Anderson localization near the transition point $W_c=8.25$. Near the critical disorder strength ($W=8\sim W_c=8.25$), the mobility edge is extended in a broad range of the energy scale ($7.6<E<8.8$) and the multifractal scaling exponent $\alpha_2$ of all electrons approaches $\alpha_2 \approx 1.13$. We also point out that the multifractal scaling exponent does not depend on the disorder strength $W$ at the mobility edge.
\vspace{-0.3cm}
\subsection{Emergence of two types of mobility edges and their multifractal scaling exponents in the presence of Coulomb interactions}
Now, we discuss various multifractal scaling exponents of $\alpha_1$, $\alpha_2$, $\alpha_{-0.5}$ and $\alpha_{-1.5}$ for the interacting case $U=0.8$ and the disorder strength $W=7$, shown in Fig.~\ref{alpha_qvar}. In addition to the mobility edge at $E_{m}^{UV}=8.91$ near the UV cutoff, an additional mobility edge appears near the band center at $E_{m}^{IR}=0.12$. The energy $E_{m}^{IR}=0.12$ corresponds to the crossover point above which the $\sim E^{2}$ behavior of the Coulomb gap in the density of states switches to the Altshuer-Aronov behavior $\sim \sqrt{E}$. See Fig.~\ref{dos1}. The multifractal scaling exponents at $E_{m}^{UV}=8.91$ near the UV cutoff turn out to be identical to the ones at the mobility edge at $E_m=8.45$ in the non-interacting case. See Fig.~\ref{alpha_U0_W7}. On the other hand, the multifractal scaling exponents at $E_{m}^{IR}=0.12$ near the Fermi energy become rather modified than the non-interacting ones at $E_{m}^{UV}=8.91$, enhanced by the factor of $1.24$ and $1.32$ for $\alpha_1$ and $\alpha_2$, respectively, and reduced by the factor of $1.24$ and $1.22$ for $\alpha_{-1/2}$ and $\alpha_{-3/2}$, respectively.

\begin{figure}
\begin{center}
  \vspace{0.2cm}
  \begin{tabular}{cc}
    \includegraphics[scale=0.25]{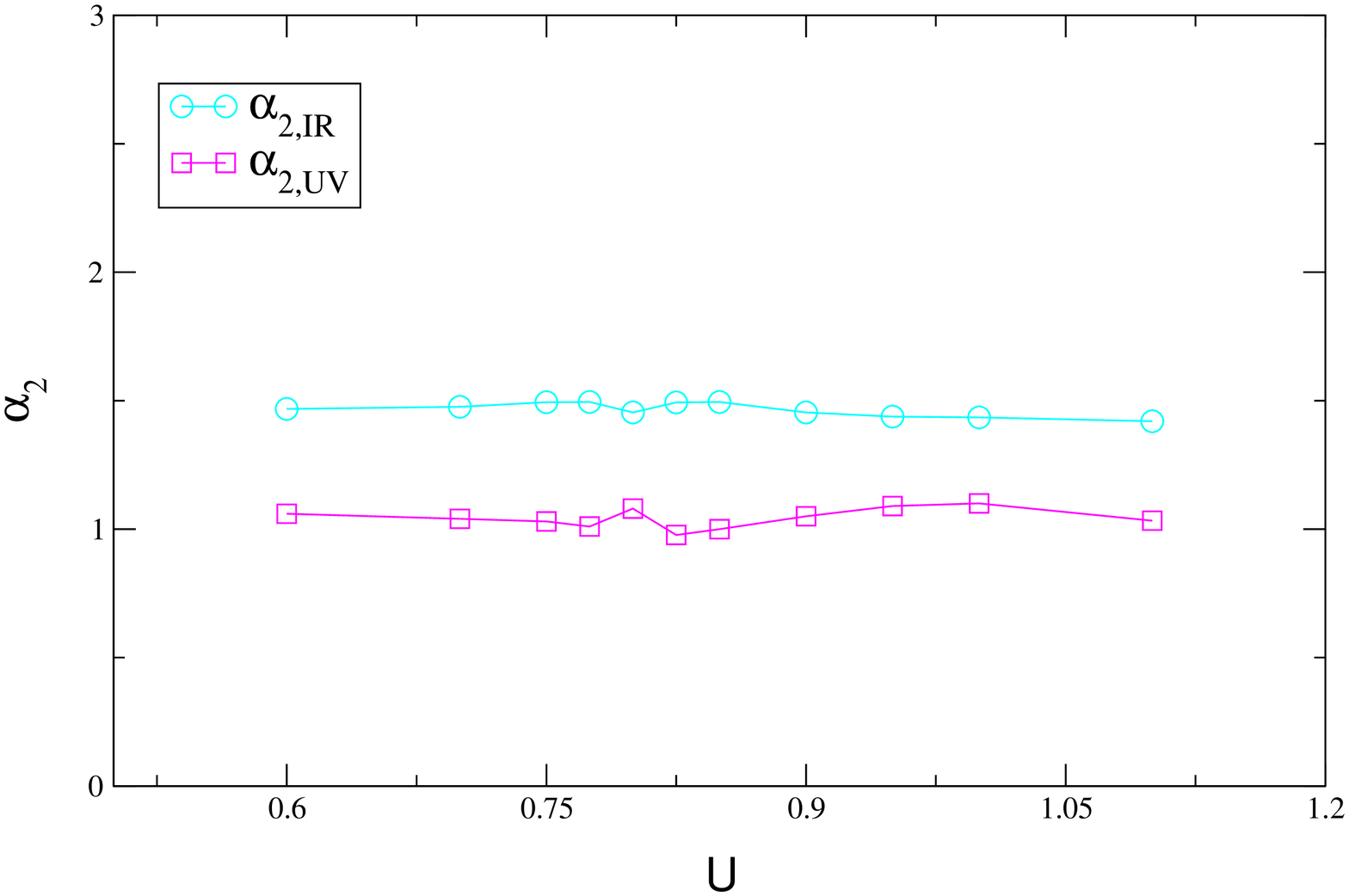}
    \vspace{-0.3cm}
    \end{tabular}
\end{center}
\caption{Multifractal scaling exponent $\alpha_2$ at the high- and the low-energy mobility edge for various interaction strengths of $0.6<U<1.1$, denoted as $\alpha_{2,UV}$ and $\alpha_{2,IR}$, respectively. The multifractal scaling exponent not only at the high-energy mobility edge but also at the low-energy mobility edge does not depend on the Coulomb interaction.}
\vspace{-0.3cm}
\label{alpha2}
\end{figure}

Fig.~\ref{alpha_Uvar} shows the multifractal scaling exponent $\alpha_2$ for four different interaction strengths $U$ and the fixed disorder strength $W=7$. The mobility edge close to the high energy cutoff remains the same as the non-interacting case. On the other hand, the mobility edge near the Fermi energy changes its position with increasing $U$ such as $E_{m}^{IR} = 0.08,~0.11,~0.12,~0.18$ for $U=0.6,~0.7,~0.8,~0.9$, respectively. The value of the critical exponent $\alpha_2$ at $E_{m}^{IR}$, however, does not depend on the strength of $U$ itself. See Fig. \ref{alpha2}.

\begin{figure}
\begin{center}
  \vspace{-0.1cm}
  \begin{tabular}{c}
    \includegraphics[scale=0.3]{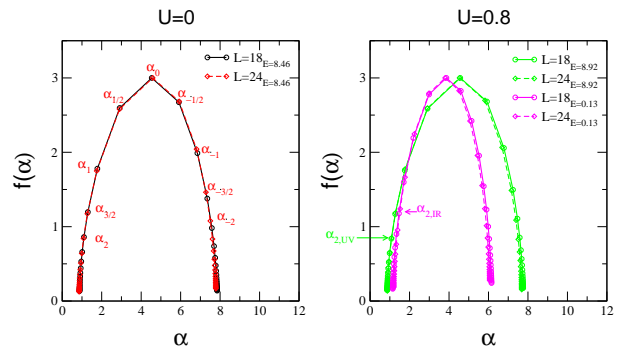}
    \vspace{-0.3cm}
    \end{tabular}
\end{center}
\caption{Multifractality singular spectrum for both noninteracting $U=0$ and interacting $U = 0.8$ cases at the disorder strength $W=7$. Regardless of the system size, the multifractality singular spectrum collapses into a single curve in the noninteracting case. On the other hand, the singularity spectrum becomes classified into two categories, which suggests the existence of two types of mobility edges.}
\vspace{-0.3cm}
\label{mult_spct}
\end{figure}

We also calculate the multifractality singular spectrum for both noninteracting $U=0$ and interacting $U = 0.8$ cases at the disorder strength $W=7$, shown in Fig.~\ref{mult_spct}. This multifractality singular spectrum contains the information of the scale invariance, and thus it does not depend on the size of a system \cite{AMIT_Review}. The left panel displays that the multifractality singular spectrum collapses into a single curve, regardless of the system size, when electron correlations are turned off. The right panel shows that the multifractality singular spectrum can be classified into two categories, corresponding to the high-energy and low-energy mobility edges, respectively. The green single curve represents the multifractality singular spectrum at the high-energy mobility edge, essentially the same as that of the left panel, and the magenta single curve does it at the low-energy mobility edge, distinguished from the noninteracting multifractal spectrum.
\vspace{-0.3cm}
\subsection{Comparison with recent analytical and numerical studies}
It is necessary to compare our numerical results with recent analytical and numerical studies. Although we focus on $\alpha_{2}$ in the present study, we also find $\tau_{2} \approx 1.7$, resorting to Eq.~(\ref{tau_q2}), where a typical value has been considered. A recent nonlinear $\sigma-$model study \cite{Multifractality_Interaction_NLsM} investigated the scaling behavior of moments of the local density of states for the unitary ensemble in the presence of Coulomb interactions, given by $[\rho(\omega,\bm{r}) / \langle \rho(\omega) \rangle]^{q}$. Based on the $\epsilon-$expansion near the lower critical dimension with $d = 2 + \epsilon$, this study found an anomalous fractal exponent $\Delta_{q} = - \frac{q(q-1) \epsilon}{4} \Big\{ 1 + \Big( 1 - A - \frac{\pi^{2}}{12} \Big) \epsilon \Big\} + \mathcal{O}(\epsilon^{3})$ up to the two-loop order, where the anomalous fractal exponent is given by $\Delta_{q} = \tau_{q} - d (q-1)$. Here, $A \approx 1.64$ is a positive numerical constant, which appears in the renormalization group equation for the inverse of the dimensionless conductance. Actually, this analytic study reported $\tau_{2} \approx 2.5$, which deviates from a recent numerical study \cite{Multifractality_Interaction_Numerics}. On the other hand, the nonlinear $\sigma-$model field theory gives rise to $\tau_{2} \approx 1.6$ up to the four-loop level in the absence of electron correlations, consistent with numerical results \cite{Multifractality_Interaction_NLsM}.

A recent Hartree-Fock numerical study reported $\tau_{2} = 1.57 \pm 0.05$ for the system size of $L = 10$ while $\tau_{2} = 1.34 \pm 0.05$ in the absence of electron interactions \cite{Multifractality_Interaction_Numerics}. This value is slightly smaller than the present typical value $\tau_{2} \approx 1.7$. Generally speaking, the fractal dimension increases in the presence of electron correlations, implying more sparse distributions of moments of eigenfunctions. This enhancement results from the fact that electron interactions give rise to linearly superposed states of fractal eigenfunctions in the absence of interactions, which weakens the multifractal nature of the Anderson metal-insulator transition. The Ewald summation technique seems to take into account long-ranged Coulomb interactions more strongly.
\section{Summary and Discussion}
\vspace{-0.3cm}

In summary, we investigated the role of Coulomb interactions in the nature of eigenfunction multifractality of an Anderson metal-insulator transition, based on the Hartree-Fock approximation and the Ewald summation technique. As a result, we showed that two types of mobility edges appear near the Fermi energy and at a high energy, respectively, where the low-energy mobility edge results from Coulomb interactions while the high-energy one is nothing but the mobility edge of the Anderson localization transition without electron correlations. Indeed, not only multifractal scaling exponents but also the multifractal singularity spectrum confirms the existence of two kinds of mobility edges: Their values differ from those of the Anderson metal-insulator transition and the singularity spectrum collapses into two types of curves, implying two kinds of scale-invariance, which depends on the energy scale. We speculate that this novel nature of the eigenfunction multifractality would serve as valuable information for possible instabilities near a metal-insulator transition in the presence of Coulomb interactions \cite{Kondo_Anderson_Transition,BCS_Anderson_Transition,Stoner_Anderson_Transition}.

Before closing, we would like to point out that our Hartree-Fock self-consistent equations with the Ewald summation technique do not take into account screening of the Coulomb interaction. In particular, the Coulomb interaction should be screened by particle-hole excitations near a Fermi surface in a metallic phase, described by the random phase approximation (RPA). Here, the RPA correction can be taken into account in a fashion of real space, given by matrix products to describe a convolution integral. Even if such corrections are not introduced into the self-consistent equations, order parameters protect the correct physics of the Altshuler-Aronov correction in a metallic state. In other words, the exchange hopping order parameter, which becomes long ranged potentially by Coulomb interactions, remains short ranged in a metallic phase, keeping the Altshuler-Aronov correction described by the Hartree-Fock approximation in the presence of disorder scattering. In an insulating phase, the Coulomb interaction itself persists, well described by the Ewald summation technique. However, the absence of the RPA correction may be dangerous in the vicinity of the metal-insulator transition because the screening effect can cause anomalous scaling behavior for the Coulomb interaction instead of the $\sim 1/ r$ potential. If this is the case, the present calculations would have uncertainties for multifractal scaling exponents. However, we emphasize that the existence of two kinds of mobility edges will not be affected by this approximation scheme, where the low energy mobility edge occurs after the formation of the Coulomb gap, i.e., in the insulating state.
\section{Acknowledgement}
This study was supported by the Ministry of Education, Science, and Technology (No. NRF-2015R1C1A1A01051629 and No. 2011-0030046) of the National Research Foundation of Korea (NRF). Computing resources were provided by the NSF via grant MRI: Acquisition of Conflux, A Novel Platform for Data-Driven Computational Physics (Tech. Monitor: Ed Walker). We appreciate helpful discussions with S. Kettemann, X. Wan, R. Narayanan, and V. Dobrosavljevic.
\section{Appendix}
\subsection{Ewald summation} \label{Ewald}
Consider $N$ charged particles subjected to the periodic boundary condition,
\begin{equation}
\rho{({\bf r})}=\rho({\bf r}+{\bf n} L),
\label{pbc}
\end{equation}
where ${\bf n}=n_1\hat{x}+n_2\hat{y}+n_3\hat{z}$ with arbitrary integers $n_1$, $n_2$ and $n_3$.
The total Coulomb interaction energy includes interactions between real and image charges in periodic supercells, given by
\begin{equation}
  E=\frac{1}{4\pi\varepsilon_0}\frac{1}{2}\sum_{\bf n}\sum_{i=1}^N{\sum_{j=1}^N}^\prime\frac{q_i q_j}{|{\bf r}_{ij}+{\bf n} L|},
\label{LR_int}
  \end{equation}
where ${\bf {r}}_{ij}+{\bf n}L$  is the distance between the two point charges $q_i$ and $q_j$ located in two separate supercells. The $\prime$ symbol means that the term $j=i$ is excluded, if and only if ${\bf n=0}$.

  In the Ewald technique, the long-range interaction in Eq.~(\ref{LR_int}) is split into two parts; a short-range interaction potential whose pairwise sum readily converges in real space and a long-range portion based on a slowly-varying interaction potential whose pairwise sum converges relatively quickly in reciprocal space \cite{Ewald_Sum_Technique1,Ewald_Sum_Technique2,Ewald_Sum_Technique3}.

The original charge distribution $\rho({\bf r})$ can be split into two terms,
\begin{eqnarray}
  \rho_i({\bf r})&=&\rho^S_i({\bf r})+\rho^L_i({\bf r}),\\
  \rho^S_i({\bf r})&=&q_i \delta({\bf r-r_i})-q_i\mathcal{G}({\bf r-r_i}), \\
  \rho^L_i({\bf r})&=&q_i\mathcal{G}({\bf r-r_i}).
  \end{eqnarray}
where $\mathcal{G}({\bf r})$ is a Gaussian distribution,
\begin{equation}
  \mathcal{G}({\bf r})=\frac{1}{(2\pi\sigma^2)^{3/2}}{\rm exp}\left[ -\frac{|{\bf r}|^2}{2\sigma^2}\right].
\end{equation}
The potential field generated by a charge distribution of the Gaussian form is obtained as
\begin{eqnarray}
%  \nabla^2 \phi_i({\bf r})&=&-\frac{1}{\varepsilon_0}\mathcal{G}({\bf r}) \\
  \phi(r) &=&\frac{1}{4\pi\varepsilon_0 r} {\rm erf}(\frac{r}{\sqrt{2}\sigma}) ,
\end{eqnarray}
where ${\rm erf}(z)\equiv \frac{2}{\sqrt{\pi}}\int_0^z e^{-t^2}dt$. Accordingly, the Coulomb potential is written as
\begin{eqnarray}
  \phi_i({\bf r})&=&\phi^S_i({\bf r})+\phi^L_i({\bf r}) , \\
  \phi^S_i({\bf r})&=&\frac{q_i}{4\pi\varepsilon_0}\frac{1}{|\bf r-r_i|}\left[{\rm erfc}\left(\frac{|{\bf r-r_i}|}{\sqrt{2}\sigma}\right)\right] ,
\label{phi_s}
\\
  \phi^L_i({\bf r})&=&\frac{q_i}{4\pi\varepsilon_0}\frac{1}{|\bf r-r_i|}\left[{\rm erf}\left(\frac{|{\bf r-r_i}|}{\sqrt{2}\sigma}\right)\right] .
  \label{phi_r}
  \end{eqnarray}
  Note that ${\rm erfc } (x)=1-{\rm erf}(x)$. Here, the Ewald parameter $\sigma$ is the cutoff length scale on which the short-range function $\frac{1}{r}{\rm erfc}(\frac{r}{\sqrt{2}\sigma})$ decays. As $\sigma$ decreases, more of the summation is performed in reciprocal space, whereas, setting $\sigma \sim L$, the Coulomb interaction is taken into account entirely in real space.

Using Eq.~(\ref{phi_s}) and Eq.~(\ref{phi_r}), the total Coulomb interaction energy in Eq.~(\ref{LR_int}) can be written as
\begin{eqnarray}
  E   &=&\frac{1}{4\pi\varepsilon_0}\frac{1}{2}\sum_{\bf n}\sum_{i=1}^N{\sum_{j=1}^N}^\prime\frac{q_i q_j}{{|\bf r_i-r_j}+{\bf n}L|}\times  \nonumber\\
 && \left[   {\rm erfc}     \left(   \frac{|{\bf r_i-r_j}+{\bf n}L|}{\sqrt{2}\sigma}   \right) +{\rm erf}\left(\frac{|{\bf r_i-r_j}+{\bf n}L|}{\sqrt{2}\sigma}\right)\right].\nonumber\\
\label{LR_int1}
\end{eqnarray}

  Now we define a cavity field $\phi_{\left[i \right]}({\bf r})$ as the potential field generated by all the ions plus their images, excluding the ion at $i$,
  \begin{equation}
 \phi_{\left[i \right]}({\bf r}) \equiv \phi({\bf r}) - \phi_{i }({\bf r}) =\frac{1}{4\pi\varepsilon_0}\sum_{\bf n}{\sum_{j=1}^{N}}^\prime \frac{q_j}{|{\bf r-r_j+n}{ L}|} .
\label{cavity}
 \end{equation}
  The $\prime$ symbol means that the term $j=i$ is excluded, if and only if ${\bf n=0}$. Using Eq.~(\ref{cavity}), the total Coulomb interaction energy in Eq.~(\ref{LR_int1}) can be written as
  \begin{eqnarray}
    E&=&\frac{1}{2} \sum_{i=1}^N q_i \phi^S_{\left[ i \right]}({\bf r}_i) +\frac{1}{2} \sum_{i=1}^N q_i \phi^L({\bf r}_i)-\frac{1}{2} \sum_{i=1}^N q_i \phi_i^L({\bf r}_i)\nonumber \\
    &=&E^S+E^L+E^{self} .
  \end{eqnarray}

With taking the $r \rightarrow 0$ limit,
\begin{eqnarray}
  \lim_{\bf r\rightarrow r_i}\phi^L_i({\bf r})&=&\frac{q_i}{4\pi\varepsilon_0}\lim_{r\rightarrow 0}\frac{1}{{r}}{\rm erf}\left(\frac{r}{\sqrt{2}\sigma}\right) ,
\label{phi_self}
\end{eqnarray}
we can easily obtain the self-energy term,
\begin{equation}
  E^{self}=\frac{1}{4\pi\sigma \varepsilon_0}\sqrt{\frac{1}{2\pi}}\frac{1}{\sigma}\sum_{i=1}^{N}q_i^2 .
\end{equation}

In order to handle the long-range portion $E^L$ in the reciprocal space, we make the Fourier transform of the total charge density
\begin{eqnarray}
  \rho^L({\bf r})&=&\sum_{\bf n}\sum_{j=1}^{N}q_j \mathcal{G}({\bf r-r_j+n}L) ,
  \end{eqnarray}
and obtain
\begin{eqnarray}
  \rho^{L}({\bf k})  &=&\mathcal{N}_{cell}\sum_{j=1}^{N} q_j e^{-i{\bf k}\cdot {\bf r_j}}e^{-\sigma^2k^2/2} ,
  \label{rho_k}
  \end{eqnarray}
where $\mathcal{N}_{cell}$ is the number of supercells. The Poisson's equation
\begin{equation}
  \nabla^2 \phi^L({\bf r})=-\frac{1}{\varepsilon_0}\rho^L({\bf r})
\end{equation}
can be Fourier-transformed into the reciprocal space, given by
\begin{equation}
  \phi^L({\bf k})=\frac{1}{\varepsilon_0}\frac{\rho^L({\bf k})}{k^2} .
  \label{poisson_k}
\end{equation}
As a result, Eq.~(\ref{rho_k}) and Eq.~(\ref{poisson_k}) give the potential field in the reciprocal space as follows
\begin{eqnarray}
  \phi^{L}({\bf k})&=&\mathcal{N}_{cell}\sum_{j=1}^{N}q_j  e^{-i{\bf k}\cdot {\bf r_j}}\frac{e^{-\sigma^2k^2/2}}{k^2}.
  \label{phi_k}
\end{eqnarray}
Applying the inverse Fourier transform, we get
\begin{eqnarray}
  \phi^{L}{(\bf r})&=&\frac{1}{V}\sum_{\bf k\neq 0} \phi^L({\bf k}) e^{i{\bf k}\cdot {\bf r}}  \nonumber\\
  &=&\frac{1}{v\varepsilon_0}\sum_{\bf k\neq 0} S({\bf k}) e^{i{\bf k}\cdot {\bf r}}\frac{e^{-\sigma^2k^2/2}}{k^2} ,
  \label{phi_L}
\end{eqnarray}
where $S({\bf k})=\sum_{j=1}^{N} q_j  e^{-i{\bf k}\cdot {\bf r_j}}$. Here, $v=\frac{V}{\mathcal{N}_{cell}}$ is the volume of a single supercell. The contribution to the ${\bf k=0}$ term is zero if the supercell is charge neutral, i.e. $\sum_i^N q_i=0.$

 In practice, we introduce an IR momentum cutoff $\varepsilon$ to neglect the small momentum contribution $|{\bf k}|<\varepsilon$ so that we can bypass the poor resolution at $|{\bf k}|<\varepsilon$ associated to the finite size of system $N=L^3$, i.e.
\begin{eqnarray}
  \phi^{L}({\bf r})  &=&\frac{1}{v\varepsilon_0}\sum_{|{\bf k}|>\varepsilon} S({\bf k}) e^{i{\bf k}\cdot {\bf r}}\frac{e^{-\sigma^2k^2/2}}{k^2}.
  \label{phi_L2}
\end{eqnarray}
Technically, the IR cutoff $\varepsilon>0$ can be regarded as an effective convergence factor which helps the summation in Eq.~(\ref{phi_L2}) absolutely convergent. Otherwise the pairwise sum of the long-range potential in Eq.~(\ref{phi_L}), which is conditionally convergent but not absolutely convergent, yields discrepant results depending on the sequence of the summation \cite{Ewald_Sum_Technique1,Ewald_Sum_Technique2}. We find that the ill-convergence of the long-range potential $\phi^L{(\bf r})$ can lead to multiplicity of Hartree-Fock solutions.

Using the results in Eq.~(\ref{phi_self}) and Eq.~(\ref{phi_L2}), the cavity potential field generated by the surrounding electrical charges can be written as
  \begin{eqnarray}
    \phi_{\left[i \right]}({\bf r_i})&=& \phi^S_{\left[i \right]}({\bf r_i})+\phi^L({\bf r_i})-\phi^{self}({\bf r_i})\nonumber \\
    &=&\frac{1}{4\pi\varepsilon_0}\sum_{\bf n}{\sum_{j=1}^{N}}^\prime q_j\frac{ {\rm erfc}\left(\frac{|{\bf r_i-r_j+n} L|}{\sqrt{2}\sigma} \right)}{|{\bf r_i-r_j+n} L|}\nonumber \\
    &+&\frac{1}{v\varepsilon_0}\sum_{|{\bf k}|>\varepsilon }\sum_{j=1}^{N} q_j  e^{i{\bf k}\cdot {\bf (r_i-r_j)}}\frac{e^{-\sigma^2k^2/2}}{k^2}\nonumber\\
    &-& \frac{q_i}{4\pi\varepsilon_0} \sqrt{\frac{2}{\pi}}\frac{1}{\sigma} .
    \label{tot_pot}
  \end{eqnarray}
Accordingly, the Hartree potential in Eq.~(\ref{selfconsol1}) is written as
\begin{eqnarray}
  \tilde{V}_i&=&\varepsilon_i-\mu +U\sum_{\bf n}{\sum_{j=1}^{N}}^\prime \delta n_j\frac{ {\rm erfc}\left(\frac{|{\bf r_i-r_j+n} L|}{\sqrt{2}\sigma} \right)}{|{\bf r_i-r_j+n} L|}\nonumber\\
  &+&\frac{4\pi U}{v}\sum_{|{\bf k}|>\varepsilon}\sum_{j=1}^{N} \delta n_j  e^{-i{\bf k}\cdot {\bf (r_i-r_j)}}\frac{e^{-\sigma^2k^2/2}}{k^2}-U\delta n_i \sqrt{\frac{2}{\pi}}\frac{1}{\sigma},\nonumber\\
  \label{tot_hartree_pot}
\end{eqnarray}
where $\delta n_i=n_i-K$ is the fluctuation of the electron occupation $n_i$ around the mean value $K$.
\begin{figure}
\begin{center}
  \vspace{0.cm}
  \begin{tabular}{c}
    \includegraphics[scale=0.31]{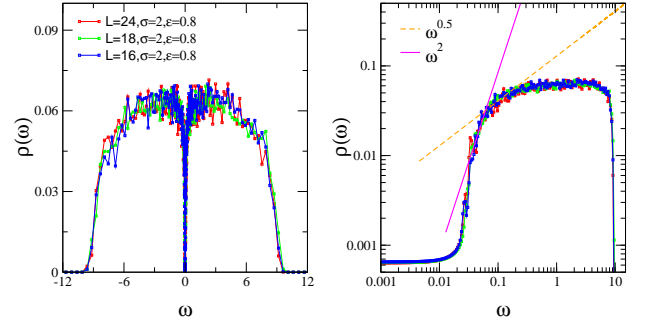}
    \end{tabular}
\end{center}
\vspace{-0.5cm}
\caption{Density of states at the interaction strength $U=0.5$ and the disorder strength $W=7$ for three different sizes of systems $N=L^3=16^3,~18^3,~24^3$ with the control parameters of $\sigma=2$ and $\varepsilon=0.8$.}
\label{dos5}
\vspace{-0.4cm}
\end{figure}

In thermodynamic limit, $N\rightarrow \infty$ and $\varepsilon \rightarrow 0$, the last two terms in Eq.~(\ref{tot_hartree_pot}) perfectly compensate each other to yield the correct power-law feature of the density of states near the Fermi level. In the presence of a finite size effect, however, the contribution of self-energy $\phi^{self}({\bf r_i})$ dominates the long-range term $\phi^L({\bf r_i})$ to open a hard gap in the density of states near zero frequency.

The parameters $\sigma$ and $\varepsilon$ in Eq.~(\ref{tot_hartree_pot}), therefore, are optimized to fulfill the two requirements. First, it should give a unique solution which is absolutely convergent, i.e., independent of the size of a system. Second, the solution should exhibit the correct power-law behavior of the density of states.

Fig.~\ref{dos5} shows the density of states evaluated at the interaction strength $U=0.5$ and the disorder strength $W=7$ for three different sizes of systems $L=16,~18$, and $24$. With $\sigma=2$ and $\varepsilon=0.8$, the density of states of these three different sizes of systems collapse into a single curve showing an insulating behavior ($\sim E^2$). The small energy region ($E<0.03$) subjected to an exponential decay is due to the mismatch between the self-energy and the long-range potential as mentioned before.
\begin{figure}
\begin{center}
  \vspace{0.cm}
  \begin{tabular}{c}
    \includegraphics[scale=0.31]{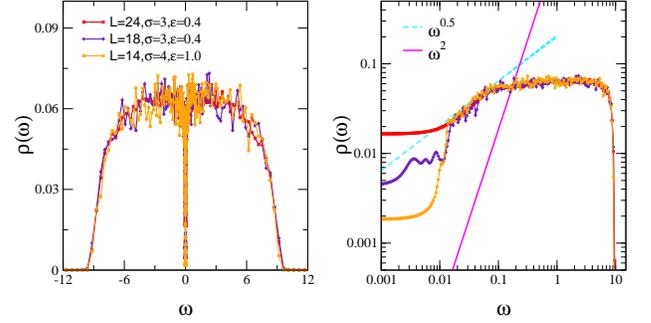}
    \end{tabular}
\end{center}
\vspace{-0.5cm}
\caption{Density of states at the interaction strength $U=0.3$ and the disorder strength $W=7$ for three different sizes of systems $N=L^3=14^3,~18^3,~24^3$  with the control parameters of $\sigma=3,~\varepsilon=0.4$ and $\sigma=4,~\varepsilon=1$.}
\label{dos6}
\vspace{-0.4cm}
\end{figure}

Fig.~\ref{dos6} shows the density of states at $U=0.3$ and $W=7$ for the size of systems $L=14,~18,~24$, where two sets of the control parameters $\sigma=3,~\varepsilon=0.4$ and $\sigma=4,~\varepsilon=1$ are considered. At present, it is not conclusive that $U=0.3$ is metallic. These three curves show the same Altshuler-Aronov behavior at the energy range $0.01<E<0.1$ but deviate from each other at $E<0.01$. If these three lines exhibit the same $\sim E^2$ below $E<0.01$ with an alternative set of parameters $\sigma$ and $\varepsilon$, the $U=0.3$ case can also correspond to an insulating phase. We find that, within the current Ewald scheme, it is more difficult to get convergence among systems with different sizes in a metallic phase.

\begin{figure}
\begin{center}
  \vspace{-0.3cm}
  \begin{tabular}{c}
    \includegraphics[scale=0.25]{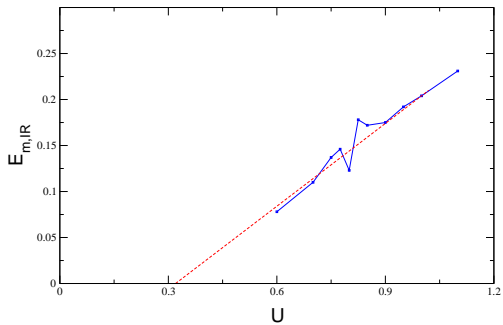}
    \end{tabular}
\end{center}
\vspace{-0.5cm}
\caption{Low-energy mobility edge $E_{m}^{IR}$ for various interaction strengths $0.5< U < 1.2$ obtained from the multifractal scaling analysis. The dashed line is linear extrapolation.}
\label{IRedge}
\vspace{-0.3cm}
\end{figure}

Fig.~\ref{IRedge} shows the low-energy mobility edge $E_{m}^{IR}$ for various interaction strengths. For $U < 0.6$, the mobility edge has an order of magnitude $10^{-2}$ and is susceptible to numerical uncertainty attributed to finite system size and the number of disorder realizations. The linear extrapolation estimates that $E_m^{IR}$ goes to zero around $U \sim 0.32$, indicating the possibility of the metal-insulator transition at $U\sim 0.3$. It is desirable to perform multifractal finite-size scaling analysis \cite{AMIT_Critical_Disorder_Strength,MFM_Numerics}, which permits the systematic analysis of the interacting mobility edge near the Fermi level.
\subsection{Long-range hopping matrix element}
  In the Hartree-Fock approximation, the hopping matrix element is self-consistently determined such as
  \begin{equation}
    \tilde{t}_{ij}=t_{ij}+\frac{U}{\bf |{r_i-r_j}|}\langle c_j^\dagger c_i\rangle
  \end{equation}
  and the effective hopping Hamiltonian is
  \begin{equation}
    \hat{H}_{hop}=\sum_{i}^{N}\sum_{j}^{N}\tilde{t}_{ij}  ({c_i}^\dagger c_j+{c_j}^\dagger c_i).
  \end{equation}
    Now consider $N$ charged particles subjected to the periodic boundary condition in Eq.~(\ref{pbc}). Including the effect of hopping between the image charges in the periodic supercells, the hopping Hamiltonian can be written as
  \begin{eqnarray}
    \hat{H}_{hop}&=&\sum^{\mathcal{N}_{cell}}_{\bf n}\sum^{\mathcal{N}_{cell}}_{\bf m} \sum_{i}^{N}\sum_{j}^{N}\tilde{t}^{\bf (nm)}_{ij}  ({c^{\bf n}_i}^\dagger c^{\bf m}_j+{c^{\bf m}_j}^\dagger c^{\bf n}_i) .
%    &=&\sum_{ij}\tilde{t}^{(\bf 00)}_{ij}({c^{\bf 0}_i}^\dagger c^{\bf 0}_j+{c^{\bf 0}_j}^\dagger c^{\bf 0}_i)\nonumber\\
%    &+&\sum_{\bf n\neq0 }\sum_{ij}\tilde{t}^{(\bf n0)}_{ij} ({c^{\bf n}_i}^\dagger c^{\bf 0}_j+{c^{\bf 0}_j}^\dagger c^{\bf n}_i)\nonumber \\
%    &+&\sum_{\bf m\neq0 }\sum_{ij}\tilde{t}^{(\bf 0m)}_{ij} ({c^{\bf 0}_i}^\dagger c^{\bf m}_j+{c^{\bf m}_j}^\dagger c^{\bf 0}_i)\nonumber\\
%    &+&\sum_{\bf n\neq0 }\sum_{\bf m\neq 0}\sum_{ij}\tilde{t}^{(\bf nm)}_{ij} ({c^{\bf n}_i}^\dagger c^{\bf m}_j+{c^{\bf m}_j}^\dagger c^{\bf n}_i).
    \end{eqnarray}
Here $\mathcal{N}_{cell}$ is the total number of cells in the supercell structure. In this work, we keep the hopping matrix elements within each supercell neglecting the intercell matrix elements,
   \begin{eqnarray}
    \hat{H}_{hop}    &\approx&\sum_{\bf n}\sum_{ij}\tilde{t}^{(\bf nn)}_{ij}({c^{\bf n}_i}^\dagger c^{\bf n}_j+{c^{\bf n}_j}^\dagger c^{\bf n}_i).
   \end{eqnarray}
   This approximation is valid if the range of electron hopping is shorter than the linear size of a cell $L$, which is easily fulfilled in the metallic or the insulating phase where the electron hopping remains short ranged. In the vicinity of the critical region, however, the hopping can also have a long-range nature and the approximation is valid only when the cell-size is large enough to cover the range of hopping.


\begin{thebibliography}{9}
\bibitem{AMIT_Review} F. Evers and A. D. Mirlin, Rev. Mod. Phys. {\bf 80}, 1355 (2008).
\bibitem{Comments_Multifractality} Multifractality implies the existence of infinitely many relevant operators, which cannot be the case for conventional continuous phase transitions. This peculiar feature may be involved with the fact that the upper critical dimension of the Anderson metal-insulator transition is infinite, where the conventional dimensional regularization technique does not work for the problem of Anderson localization.
\bibitem{Multifractality_NLsM_vs_Numerics} F. Evers and A. D. Mirlin Phys. Rev. Lett. {\bf 84}, 3690 (2000).
\bibitem{Multifractality_Interaction_Experiment} A. Richardella et al., Science {\bf 327}, 665 (2010).
\bibitem{Multifractality_Interaction_NLsM} I. S. Burmistrov, I. V. Gornyi, and A. D. Mirlin, Phys. Rev. Lett. {\bf 111}, 066601 (2013).
\bibitem{Multifractality_Interaction_Numerics} M. Amini, V. E. Kravtsov, and M. Muller, New J. Phys. {\bf 16}, 015022 (2014).
\bibitem{Ewald_Sum_Technique1} P. P. Ewald, Ann. Phys. (Leipzig) {\bf 64}, 253 (1921)
\bibitem{Ewald_Sum_Technique2} S. W. de Leeuw, J. W. Perram and E. R. Smith, Proc. Roy. Soc. Lond. A {\bf 373}, 27-56 (1980); ibid. Proc. Roy. Soc. Lond. A {\bf 373}, 57 (1980).
\bibitem{Ewald_Sum_Technique3}  H. Lee and W. Cai,  {\it Ewald summation for Coulomb interactions in a periodic supercell.} (Lecture Notes, Stanford University, 2009)
\bibitem{Anderson_Localization_Review} P. A. Lee and T. V. Ramakrishnan, Rev. Mod. Phys. {\bf 57}, 287 (1985).
\bibitem{Altshuer_Aronov_Correction} B. L. Altshuer, A. G. Aronov, A. L. Efros, and M. Pollak, \textit{Electron-electron Interactions in Disordered Systems} (Elsevier, Amsterdam, 1985)
\bibitem{Coulomb_Gap_Review} B. I. Shklovskii and A. L. Efros, \textit{Electronic properties of doped semiconductors} {\bf 45} (Springer Science $\&$ Business Media, 2013).
\bibitem{Coulomb_Disorder_DFT_I} Y. Harashima and K. Slevin, Phys. Rev. B {\bf 89}, 205108 (2014).
\bibitem{Coulomb_Disorder_DFT_II} Edoardo G. Carnio, Nicholas D. M. Hine, and Rudolf A. Romer, arXiv:1710.01742 [cond-mat.dis-nn].
\bibitem{AMIT_Critical_Disorder_Strength} K. Slevin and T. Ohtsuki, Phys. Rev. Lett. {\bf 82}, 382 (1999).
\bibitem{IRcutoff_mismatch} The exponential decay of the density of states at $\omega \sim 10^{-2}$ is exhibited even in the metallic phase $(U=0.3)$ due to the finite size of a system, which causes a mismatch between the self-energy and the long-range potential as discussed in App.~\ref{Ewald}.
\bibitem{McMillan_Shklovskii_scaling_theory} W. L. McMillan, Phys. Rev. B {\bf 24}, 2739 (1981).
\bibitem{MFM_Numerics} A. Rodriguez, Louella J. Vasquez, K. Slevin, and R. A. Roemer, Phys. Rev. B {\bf 84}, 134209 (2011); ibid. Phys. Rev. Lett. {\bf 105}, 046403 (2010).
\bibitem{f_alpha} In this work, the multifractal spectrum $f(\alpha_q)$ is directly calculated using the method proposed by Chhabra and Jensen \cite{Chhabra1989,Janssen1994},
\begin{eqnarray}
  \alpha_q&=&\frac{1}{\ln \lambda} \sum_k x_k \cdot \ln \mu_k, \\
  f(\alpha_q)&=&\frac{1}{\ln \lambda} \sum_i x_k \cdot \ln x_k ,
\end{eqnarray}
where $x_k$ is the $q$-dependent normalized quantity  $x_k={\left[ \mu_k(b)\right]^q}/{\sum_k \left[ \mu_k(b)\right]^q}$. The box probability  $\mu_k(b)$ is defined in Eq.~(\ref{boxprob}).
\bibitem{Chhabra1989}  A. Chhabra, R. V. Jensen, Phys. Rev. Lett. {\bf 62}, 1327 (1989).
\bibitem{Janssen1994}  M. Janssen, Int. J. Mod. Phys. B {\bf 8}, 943 (1994).
\bibitem{Kondo_Anderson_Transition} A. Zhuravlev, I. Zharekeshev, E. Gorelov, A. I. Lichtenstein, E. R. Mucciolo, and S. Kettemann, Phys. Rev. Lett. {\bf 99}, 247202 (2007); S. Kettemann, E. R. Mucciolo, and I. Varga, ibid. {\bf 103}, 126401 (2009); S. Kettemann, E. R. Mucciolo, I. Varga, and K. Slevin, Phys. Rev. B {\bf 85}, 115112 (2012).
\bibitem{BCS_Anderson_Transition} M. V. Feigelman, L. B. Ioffe, V. E. Kravtsov, and E. A. Yuzbashyan, Phys. Rev. Lett. {\bf 98}, 027001 (2007); M. V. Feigel¡¯man, L. B. Ioffe, V. E. Kravtsov, and E. Cuevas, Ann. Phys. {\bf 325}, 1390 (2010).
\bibitem{Stoner_Anderson_Transition} Rayda Gammag and Ki-Seok Kim, Phys. Rev. B {\bf 93}, 205128 (2016).
\end{thebibliography}
\end{document}